\documentclass[floatfix,preprint,prb,twocolumn]{revtex4}
\usepackage{graphicx}

\newcommand{\micron}{\ensuremath{\mu}\mbox{m}}

\begin{document}
\title{Polarization conversion in a silica microsphere}

\author{Pablo Bianucci}
\author{Chris Fietz}
\author{John W. Robertson}
\author{Gennady Shvets}
\author{Chih-Kang Shih}
\email{shih@physics.utexas.edu}
\affiliation{Physics Department, The University of Texas at Austin, Austin,
             Texas 78712}

\date{May 22nd, 2007}

\begin{abstract}
We experimentally demonstrate controlled polarization-selective phenomena in a
whispering gallery mode resonator.  We observed efficient ($\approx 75 \%$)
polarization conversion of light in a silica microsphere coupled to a tapered
optical fiber with proper optimization of the polarization of the propagating
light. A simple model treating the microsphere as a ring resonator provides a
good fit to the observed behavior.
\end{abstract}

\pacs{42.60.Da,42.25.Ja,42.81.Gs}

\maketitle


In the past few years, microresonators have received a lot of
attention\cite{VahalaNAT03}.  Whispering gallery mode (WGM)
resonators\cite{MatskoTQE06}, such as microspheres,\cite{GorodetskyOPL96}
microtoroids\cite{ArmaniNAT03} and microrings\cite{MelloniOPL04} have been the
object of intensive research, both in their fundamental properties (such as
quality factors, non-linear effects\cite{FominJOSAB05,CarmonPRL05} and coupling
to quantum systems\cite{ParkNL06} among many) and applications that include
lasers\cite{CaiOPL01,ShopovaAPL04}, chemical\cite{ArmaniOPL06} and
biological\cite{ArnoldOPL03} sensing and photonic
devices\cite{MichelottiAIP03}. Microsphere resonators, particularly when
coupled to a tapered optical fiber\cite{KnightOPL97,CaiPRL00}, are very useful
to characterize these properties and test new ideas due to their high Q-factors
and ease of fabrication.

Recent reports have shown a further step, taking into account the difference
between modes with different polarizations in microspheres.  In particular,
changes in the output polarization after coupling into the resonator have been
observed\cite{GuanAPL05} and transverse electric (TE) and transverse magnetic
(TM) modes have been discriminated\cite{KonishiAPL06}.

Polarization conversion has been observed in microrings\cite{MelloniOPL04} and
explained as a resonant enhancement of polarization coupling caused by waveguide
bending.  However, the mode structure of microspheres makes it possible to
completely decouple the polarizations and still obtain conversion.  In this
article, we report on the observation of efficient, controlled polarization
conversion by using a silica microsphere resonator coupled to a tapered optical
fiber. We demonstrate that highly efficient polarization conversion (75\% for
our particular case, higher for better optimized conditions) is enabled by a
specific orientation between the incoming light polarization and fiber-resonator
displacement.  Specifically, for a horizontally stacked, strongly coupled, fiber
and resonator combination, a $45^\circ$ incident polarization results in the
largest conversion.  The conversion results in a strong dip of the transmitted
light with the original polarization and a strong spike in the orthogonally
polarized transmission.
 

We fabricated the tapered fiber using the ``flame brush''
technique\cite{BirksJLT92}.  This technique involves mechanically stretching
the optical fiber while scanning a flame (oxy-hydrogen in our case) over the
region to be tapered.  Due to constraints in the maximum pulling length, the
fiber tapers are not completely adiabatic, but typical losses are never larger
than 50\%.  SEM studies of the tapers reveal a characteristic diameter close to
1 \micron.  The microsphere was fabricated using a $\mbox{CO}_2$ laser to
stretch and melt an optical fiber tip\cite{WeissOPL95}. In this way it is easy
to obtain spheres with diameters ranging from 10 \micron\ to 200 \micron. For
this particular experiment the sphere diameter was measured using an optical
microscope to be 52 \micron\ (corresponding to an estimated free spectral range
of 1.2 THz). 

We mounted the microsphere on a piezoelectric scanner which allowed us to finely
position the sphere over a range of a few micrometers, and the stretched fiber
taper on a piezoelectric stick-slip walker permitting both coarse and fine
positioning of the fiber taper next to the sphere.  Both sphere and taper were
then situated inside a compact, closed chamber.  We used an external cavity
tunable diode laser purchased from New Focus as the excitation source, centered
at a wavelength near 927.85 nm.  The polarization rotator set the polarization
of the laser which was then coupled into the optical fiber using a free-space
coupler. A polarizer and an amplified photodiode at the fiber output were used
to analyze the transmitted light. 

\begin{figure}[t]
  \centering
  \includegraphics[width=3in,keepaspectratio]{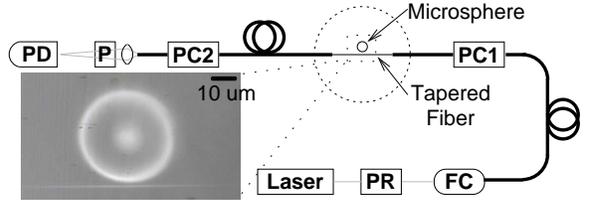}
  \caption{Experimental setup schematic. PR is a polarization rotator, FC a 
           fiber coupler, PC1 and 2 are fiber polarization controllers, P a
           polarizer and PD is an amplified photodiode. Inset: Image of a
           sphere near a tapered fiber.}
  \label{fig:setup}
\end{figure}

Space constraints in the chamber and limitations on the arrangement of the
optical fiber caused bending of the fiber in different locations and subsequent
scrambling of the input polarization.  As a way to compensate for these changes
in the polarization, we used two polarization controllers.  The first one
preceded the fiber taper, compensating for polarization changes up to the
position of the microsphere.  The second controller was placed after the fiber
taper to ensure the linearity of the output polarization.  Figure
\ref{fig:setup} shows a schematic of this experimental setup.


We used the following procedure to measure the degree of polarization
conversion.  First, the incoming polarization was selected by using the
polarization rotator.  Then we adjusted the first polarization controller to
ensure the polarization at the fiber taper was linear and matched to one set of
modes (``x-polarized'').  The next step was to uncouple the taper from the
sphere and make sure the output polarization was linear (we achieved this by
turning the detection polarizer to its position for minimum transmission and
then minimizing this transmission further with the second polarization
controller).  This orientation of the detection polarizer is the one we call
``orthogonal''.  Rotating the polarizer 90 degrees (the ``parallel''
orientation) resulted in maximum transmission, with a contrast of about 95\%,
confirming the linear polarization of the output light.  Finally, we positioned
the sphere and the tapered fiber trying to optimize the coupling, while
measuring transmission spectra for both orientations of the detection polarizer. 
We repeated the procedure for two other incoming polarizations:  one matched to
the other set of sphere modes (``y-polarized'') and another at $45^\circ$
between the x- and y- polarization axis (``xy-polarized''). 

\begin{figure}[t]
  \centering
  \includegraphics[width=3in,keepaspectratio]{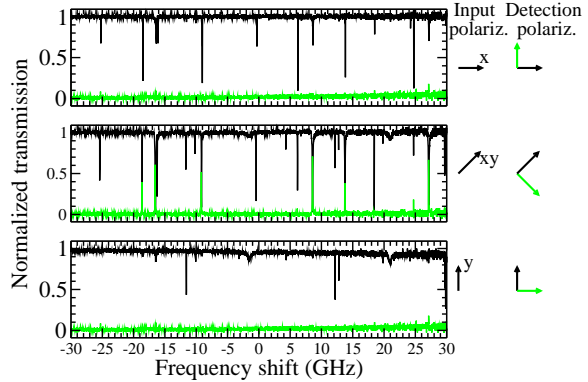}
  \caption{Transmission spectra for different input polarizations. The resonant
	   frequencies correspond to modes with $l \approx 496$. The x- and
	   y-polarizations are orthogonal and correspond to the polarization
	   eigenmodes of the resonator.  The xy-polarization is oriented at 45
	   degrees from both x and y.  The dark traces correspond to the
	   detection polarizer parallel to the input polarization and the light
	   traces correspond to a crossed detection polarization.}
  \label{fig:summary}
\end{figure}

Figure \ref{fig:summary} shows the resulting transmission spectra for the
different configurations.  The cases for both the x- and y- polarized light show
the same behavior:  a set of transmission dips whenever the laser frequency hit
a whispering gallery resonance when the detection polarization is parallel and
no signal when it is perpendicular.  The xy-polarized case is more interesting: 
the parallel detection polarization shows dips for both sets of modes, while the
orthogonal one shows transmission peaks at the whispering gallery resonances. 
At the highest peak, more than 70\% of the incident light had its polarization
converted. 

Most of the observed polarization conversion can be understood by using a simple
ring resonator model for the whispering gallery modes.  In this model, the
transmission of polarized light through the resonator is given
by\cite{SMithJOSB03,CaiPRL00}
\begin{equation}
  \tau(\phi) =
      \frac{r-ae^{i\phi}}{1-rae^{i\phi}},
  \label{eq:resonator}
\end{equation}
where $r$ is the field coupling coefficient between the resonator and the
waveguide, $a$ is the attenuation due to the resonator intrinsic losses and
$\phi = 2\pi(\nu - \nu_0)t_{RT}$ is the phase shift imposed by the resonator
($\nu$ and $\nu_0$ are the incoming light frequency and the resonant frequency
respectively, while $t_{RT}$ is the round-trip time in the resonator).  The
model is scalar, but we can include the polarization by simply assuming that
modes with orthogonal polarizations are independent and neglecting
cross-polarized couplings (using an analysis similar to that by Little and
Chu\cite{LittlePTL00}).  In this way we obtain the same expression, with
possibly different parameters, for the transmission of both polarizations.  In
our particular case of whispering gallery modes in microspheres, we can safely
assume that modes with different polarizations are not degenerate, so one of the
polarizations will be unaffected by the presence of a resonance.  This differs
from the case of microrings\cite{MelloniOPL04}, where the conversion depends on
coupling between TE and TM modes.


The essence of the effect lies in the different resonator response for each
polarization.  For a strongly coupled fiber and microsphere, $|\tau| \approx 1$,
but the phase shift $\psi = \arg(\tau)$ is changed by $\Delta\psi = \pi$ as
the frequency is sweeped across the resonance.  Because the orthogonal
polarization is transmitted unaltered, the transmitted polarization rotates by
as much as $90^\circ$ for the initial xy-polarization.  When the fiber and the
resonator are horizontally stacked, the effect is maximized when the incident
polarization is at $45^\circ$ degrees with respect to the horizontal plane.

Conversion efficiencies of up to 25\% can be achieved if one of the
polarizations is critically coupled to the ring, i.e. is completely absorbed
in+the resonator.  Achieving higher efficiencies requires increasing the
resonator-waveguide coupling to obtain a significant polarization dependent
phase shift which will change the final polarization state into one closer to
the desired one.

\begin{figure}[t]
  \centering
  \includegraphics[width=3in,keepaspectratio]{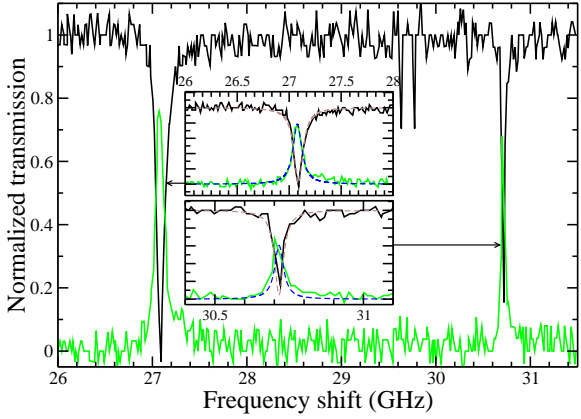}
  \caption{Detailed view of two modes showing polarization conversion.  The
	   dashed lines are fits using equations of the form of equation
	   \ref{eq:resonator}.  The fit parameters for the leftmost features are
	   $a=0.99997$, $r=0.99977$.  The corresponding ones for the rightmost
	   feature are $a=0.99999$, $r=0.99993$.}
  \label{fig:detail}
\end{figure}

We can look in more detail at the data by concentrating into a pair of modes
showing good conversion, now accounting for laser frequency drift between scans
using a Fabry-Perot interferometer as a reference.  This detailed spectrum can
be seen in in Fig.  \ref{fig:detail}.  The resonance on the right side of Fig. 
\ref{fig:detail}, near a shift of 31 GHz, shows a polarization conversion of
about 60\%.  The left-side resonance shows a conversion near 75\%.  The higher
efficiency is due to the leftmost mode being more strongly coupled (displaying a
broader feature) to the tapered fiber than the rightmost one.  Consistent with
theoretical predictions, in both cases one of the polarizations is over-coupled
to the ring.  The lack of a shift in the center frequency of the features also
indicates that each pair of peak and dip corresponds to a single resonant mode. 

\begin{figure}[t]
  \centering
  \includegraphics[width=3in,keepaspectratio]{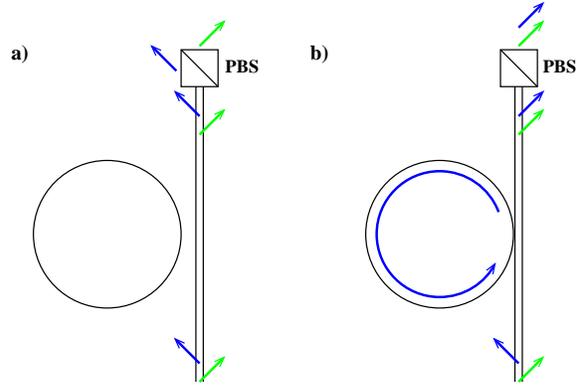}
  \caption{Schematic of a resonator working as wavelength-selective polarization
	  switch.  {\bf a)} Two signals with different wavelengths+ (green and
	  blue) and orthogonal polarizations pass unchanged through the
	  waveguide and the uncoupled resonator.  A polarization beamsplitter
	  then routes the signals to different paths.  {\bf b)} The polarization
	  of the resonant signal (blue) is converted by the coupled resonator,
	  and both signals are sent through the same path.  The
	  resonator-waveguide coupling can be changed in different ways,
	  including mechanical or optical\cite{AlmeidaNAT04} means.}
  \label{fig:switch}
\end{figure}

This phenomenon could be useful for polarization control in photonic devices,
such as narrowband polarization-dependent filtering or switching, as shown in
Fig. \ref{fig:switch} or even arbitrary polarization manipulation.

We have observed efficient polarization conversion on a microsphere resonator
coupled to a tapered optical fiber and used a simple theoretical model to
understand the phenomenon.  The model does not involve direct coupling of the
orthogonal polarizations, but rather a polarization-selective phase shift
induced by the resonator.  This effect should be common to all whispering
gallery mode resonators and could be useful for polarization control in photonic
devices.

\begin{acknowledgments}
This work was supported by NSF-NIRT (DMR-0210383), the Texas Advanced
Technology program, and the W.M. Keck Foundation.  G.S.  and C.F.  acknowledge
support from ARO MURI grant no.  W911NF-04-01-0203. 
\end{acknowledgments}



\begin{thebibliography}{22}
\expandafter\ifx\csname natexlab\endcsname\relax\def\natexlab#1{#1}\fi
\expandafter\ifx\csname bibnamefont\endcsname\relax
  \def\bibnamefont#1{#1}\fi
\expandafter\ifx\csname bibfnamefont\endcsname\relax
  \def\bibfnamefont#1{#1}\fi
\expandafter\ifx\csname citenamefont\endcsname\relax
  \def\citenamefont#1{#1}\fi
\expandafter\ifx\csname url\endcsname\relax
  \def\url#1{\texttt{#1}}\fi
\expandafter\ifx\csname urlprefix\endcsname\relax\def\urlprefix{URL }\fi
\providecommand{\bibinfo}[2]{#2}
\providecommand{\eprint}[2][]{\url{#2}}

\bibitem[{\citenamefont{Vahala}(2003)}]{VahalaNAT03}
\bibinfo{author}{\bibfnamefont{K.~J.} \bibnamefont{Vahala}},
  \bibinfo{journal}{\nat} \textbf{\bibinfo{volume}{424}}, \bibinfo{pages}{839}
  (\bibinfo{year}{2003}).

\bibitem[{\citenamefont{Matsko and Ilchenko}(2006)}]{MatskoTQE06}
\bibinfo{author}{\bibfnamefont{A.~B.} \bibnamefont{Matsko}} \bibnamefont{and}
  \bibinfo{author}{\bibfnamefont{V.~S.} \bibnamefont{Ilchenko}},
  \bibinfo{journal}{IEEE J. Sel. Top. Quantum Electron.}
  \textbf{\bibinfo{volume}{12}}, \bibinfo{pages}{3} (\bibinfo{year}{2006}).

\bibitem[{\citenamefont{Gorodetsky et~al.}(1996)\citenamefont{Gorodetsky,
  Savchenkov, and Ilchenko}}]{GorodetskyOPL96}
\bibinfo{author}{\bibfnamefont{M.~L.} \bibnamefont{Gorodetsky}},
  \bibinfo{author}{\bibfnamefont{A.~A.} \bibnamefont{Savchenkov}},
  \bibnamefont{and} \bibinfo{author}{\bibfnamefont{V.~S.}
  \bibnamefont{Ilchenko}}, \bibinfo{journal}{\ol}
  \textbf{\bibinfo{volume}{21}}, \bibinfo{pages}{453} (\bibinfo{year}{1996}).

\bibitem[{\citenamefont{Armani et~al.}(2003)\citenamefont{Armani, Kippenberg,
  Spillane, and Vahala}}]{ArmaniNAT03}
\bibinfo{author}{\bibfnamefont{D.~K.} \bibnamefont{Armani}},
  \bibinfo{author}{\bibfnamefont{T.~J.} \bibnamefont{Kippenberg}},
  \bibinfo{author}{\bibfnamefont{S.~M.} \bibnamefont{Spillane}},
  \bibnamefont{and} \bibinfo{author}{\bibfnamefont{K.~J.}
  \bibnamefont{Vahala}}, \bibinfo{journal}{\nat}
  \textbf{\bibinfo{volume}{421}}, \bibinfo{pages}{925} (\bibinfo{year}{2003}).

\bibitem[{\citenamefont{Melloni et~al.}(2004)\citenamefont{Melloni, Morichetti,
  and Martinelli}}]{MelloniOPL04}
\bibinfo{author}{\bibfnamefont{A.}~\bibnamefont{Melloni}},
  \bibinfo{author}{\bibfnamefont{F.}~\bibnamefont{Morichetti}},
  \bibnamefont{and}
  \bibinfo{author}{\bibfnamefont{M.}~\bibnamefont{Martinelli}},
  \bibinfo{journal}{\ol} \textbf{\bibinfo{volume}{29}}, \bibinfo{pages}{2785}
  (\bibinfo{year}{2004}).

\bibitem[{\citenamefont{Fomin et~al.}(2005)\citenamefont{Fomin, Gorodetsky,
  Grudinin, and Ilchenko}}]{FominJOSAB05}
\bibinfo{author}{\bibfnamefont{A.~E.} \bibnamefont{Fomin}},
  \bibinfo{author}{\bibfnamefont{M.~L.} \bibnamefont{Gorodetsky}},
  \bibinfo{author}{\bibfnamefont{I.~S.} \bibnamefont{Grudinin}},
  \bibnamefont{and} \bibinfo{author}{\bibfnamefont{V.~S.}
  \bibnamefont{Ilchenko}}, \bibinfo{journal}{\josab}
  \textbf{\bibinfo{volume}{22}}, \bibinfo{pages}{459} (\bibinfo{year}{2005}).

\bibitem[{\citenamefont{Carmon et~al.}(2005)\citenamefont{Carmon, Rokhsari,
  Yang, Kippenberg, and Vahala}}]{CarmonPRL05}
\bibinfo{author}{\bibfnamefont{T.}~\bibnamefont{Carmon}},
  \bibinfo{author}{\bibfnamefont{H.}~\bibnamefont{Rokhsari}},
  \bibinfo{author}{\bibfnamefont{L.}~\bibnamefont{Yang}},
  \bibinfo{author}{\bibfnamefont{T.}~\bibnamefont{Kippenberg}},
  \bibnamefont{and} \bibinfo{author}{\bibfnamefont{K.~J.}
  \bibnamefont{Vahala}}, \bibinfo{journal}{\prl} \textbf{\bibinfo{volume}{94}},
  \bibinfo{pages}{223902} (\bibinfo{year}{2005}).

\bibitem[{\citenamefont{Park et~al.}(2006)\citenamefont{Park, Cook, and
  Wang}}]{ParkNL06}
\bibinfo{author}{\bibfnamefont{Y.-S.} \bibnamefont{Park}},
  \bibinfo{author}{\bibfnamefont{A.~K.} \bibnamefont{Cook}}, \bibnamefont{and}
  \bibinfo{author}{\bibfnamefont{H.}~\bibnamefont{Wang}},
  \bibinfo{journal}{Nano. Lett.} \textbf{\bibinfo{volume}{6}},
  \bibinfo{pages}{2075} (\bibinfo{year}{2006}).

\bibitem[{\citenamefont{Cai and Vahala}(2001)}]{CaiOPL01}
\bibinfo{author}{\bibfnamefont{M.}~\bibnamefont{Cai}} \bibnamefont{and}
  \bibinfo{author}{\bibfnamefont{K.}~\bibnamefont{Vahala}},
  \bibinfo{journal}{\ol} \textbf{\bibinfo{volume}{26}}, \bibinfo{pages}{884}
  (\bibinfo{year}{2001}).

\bibitem[{\citenamefont{Shopova et~al.}(2004)\citenamefont{Shopova, Farca,
  Rosenberger, Wickramanayake, and Kotov}}]{ShopovaAPL04}
\bibinfo{author}{\bibfnamefont{S.~I.} \bibnamefont{Shopova}},
  \bibinfo{author}{\bibfnamefont{G.}~\bibnamefont{Farca}},
  \bibinfo{author}{\bibfnamefont{A.~T.} \bibnamefont{Rosenberger}},
  \bibinfo{author}{\bibfnamefont{W.~M.} \bibnamefont{Wickramanayake}},
  \bibnamefont{and} \bibinfo{author}{\bibfnamefont{N.~A.} \bibnamefont{Kotov}},
  \bibinfo{journal}{\apl} \textbf{\bibinfo{volume}{85}}, \bibinfo{pages}{6101}
  (\bibinfo{year}{2004}).

\bibitem[{\citenamefont{Armani and Vahala}(2006)}]{ArmaniOPL06}
\bibinfo{author}{\bibfnamefont{A.~M.} \bibnamefont{Armani}} \bibnamefont{and}
  \bibinfo{author}{\bibfnamefont{K.~J.} \bibnamefont{Vahala}},
  \bibinfo{journal}{\ol} \textbf{\bibinfo{volume}{31}}, \bibinfo{pages}{1896}
  (\bibinfo{year}{2006}).

\bibitem[{\citenamefont{Arnold et~al.}(2003)\citenamefont{Arnold, Khoshsima,
  Teraoka, Holler, and Vollmer}}]{ArnoldOPL03}
\bibinfo{author}{\bibfnamefont{S.}~\bibnamefont{Arnold}},
  \bibinfo{author}{\bibfnamefont{M.}~\bibnamefont{Khoshsima}},
  \bibinfo{author}{\bibfnamefont{I.}~\bibnamefont{Teraoka}},
  \bibinfo{author}{\bibfnamefont{S.}~\bibnamefont{Holler}}, \bibnamefont{and}
  \bibinfo{author}{\bibfnamefont{F.}~\bibnamefont{Vollmer}},
  \bibinfo{journal}{\ol} \textbf{\bibinfo{volume}{28}}, \bibinfo{pages}{272}
  (\bibinfo{year}{2003}).

\bibitem[{\citenamefont{Michelotti et~al.}(2003)\citenamefont{Michelotti,
  Driessen, and Bertolotti}}]{MichelottiAIP03}
\bibinfo{editor}{\bibfnamefont{F.}~\bibnamefont{Michelotti}},
  \bibinfo{editor}{\bibfnamefont{A.}~\bibnamefont{Driessen}}, \bibnamefont{and}
  \bibinfo{editor}{\bibfnamefont{M.}~\bibnamefont{Bertolotti}}, eds.,
  \emph{\bibinfo{title}{Microresonators as building blocks for VLSI
  photonics}}, vol. \bibinfo{volume}{709} of \emph{\bibinfo{series}{AIP
  Conference Proceedings}} (\bibinfo{publisher}{American Institute of Physics},
  \bibinfo{address}{Melville, New York}, \bibinfo{year}{2003}).

\bibitem[{\citenamefont{Cai et~al.}(2000)\citenamefont{Cai, Painter, and
  Vahala}}]{CaiPRL00}
\bibinfo{author}{\bibfnamefont{M.}~\bibnamefont{Cai}},
  \bibinfo{author}{\bibfnamefont{O.}~\bibnamefont{Painter}}, \bibnamefont{and}
  \bibinfo{author}{\bibfnamefont{K.~J.} \bibnamefont{Vahala}},
  \bibinfo{journal}{\prl} \textbf{\bibinfo{volume}{85}}, \bibinfo{pages}{74}
  (\bibinfo{year}{2000}).

\bibitem[{\citenamefont{Knight et~al.}(1997)\citenamefont{Knight, Cheung,
  Jacques, and Birks}}]{KnightOPL97}
\bibinfo{author}{\bibfnamefont{J.~C.} \bibnamefont{Knight}},
  \bibinfo{author}{\bibfnamefont{G.}~\bibnamefont{Cheung}},
  \bibinfo{author}{\bibfnamefont{F.}~\bibnamefont{Jacques}}, \bibnamefont{and}
  \bibinfo{author}{\bibfnamefont{T.~A.} \bibnamefont{Birks}},
  \bibinfo{journal}{\ol} \textbf{\bibinfo{volume}{22}}, \bibinfo{pages}{1129}
  (\bibinfo{year}{1997}).

\bibitem[{\citenamefont{Guan and Vollmer}(2005)}]{GuanAPL05}
\bibinfo{author}{\bibfnamefont{G.}~\bibnamefont{Guan}} \bibnamefont{and}
  \bibinfo{author}{\bibfnamefont{F.}~\bibnamefont{Vollmer}},
  \bibinfo{journal}{\apl} \textbf{\bibinfo{volume}{86}},
  \bibinfo{pages}{121115} (\bibinfo{year}{2005}).

\bibitem[{\citenamefont{Konishi et~al.}(2006)\citenamefont{Konishi, Fujiwara,
  Takeuchi, and Sasaki}}]{KonishiAPL06}
\bibinfo{author}{\bibfnamefont{H.}~\bibnamefont{Konishi}},
  \bibinfo{author}{\bibfnamefont{H.}~\bibnamefont{Fujiwara}},
  \bibinfo{author}{\bibfnamefont{S.}~\bibnamefont{Takeuchi}}, \bibnamefont{and}
  \bibinfo{author}{\bibfnamefont{K.}~\bibnamefont{Sasaki}},
  \bibinfo{journal}{\apl} \textbf{\bibinfo{volume}{89}},
  \bibinfo{pages}{121107} (\bibinfo{year}{2006}).

\bibitem[{\citenamefont{Birks and Li}(1992)}]{BirksJLT92}
\bibinfo{author}{\bibfnamefont{T.~A.} \bibnamefont{Birks}} \bibnamefont{and}
  \bibinfo{author}{\bibfnamefont{Y.~W.} \bibnamefont{Li}}, \bibinfo{journal}{J.
  Lightwave Technol.} \textbf{\bibinfo{volume}{10}}, \bibinfo{pages}{432}
  (\bibinfo{year}{1992}).

\bibitem[{\citenamefont{Weiss et~al.}(1995)\citenamefont{Weiss, Sandoghar,
  Hare, Lef\`{e}vre-Seguin, Raimond, and Haroche}}]{WeissOPL95}
\bibinfo{author}{\bibfnamefont{D.~S.} \bibnamefont{Weiss}},
  \bibinfo{author}{\bibfnamefont{V.}~\bibnamefont{Sandoghar}},
  \bibinfo{author}{\bibfnamefont{J.}~\bibnamefont{Hare}},
  \bibinfo{author}{\bibfnamefont{V.}~\bibnamefont{Lef\`{e}vre-Seguin}},
  \bibinfo{author}{\bibfnamefont{J.-M.} \bibnamefont{Raimond}},
  \bibnamefont{and} \bibinfo{author}{\bibfnamefont{S.}~\bibnamefont{Haroche}},
  \bibinfo{journal}{\ol} \textbf{\bibinfo{volume}{20}}, \bibinfo{pages}{1835}
  (\bibinfo{year}{1995}).

\bibitem[{\citenamefont{Smith et~al.}(2003)\citenamefont{Smith, Chang, and
  Fuller}}]{SMithJOSB03}
\bibinfo{author}{\bibfnamefont{D.~D.} \bibnamefont{Smith}},
  \bibinfo{author}{\bibfnamefont{H.}~\bibnamefont{Chang}}, \bibnamefont{and}
  \bibinfo{author}{\bibfnamefont{K.~A.} \bibnamefont{Fuller}},
  \bibinfo{journal}{\josab} \textbf{\bibinfo{volume}{20}},
  \bibinfo{pages}{1967} (\bibinfo{year}{2003}).

\bibitem[{\citenamefont{Little and Chu}(2000)}]{LittlePTL00}
\bibinfo{author}{\bibfnamefont{B.~E.} \bibnamefont{Little}} \bibnamefont{and}
  \bibinfo{author}{\bibfnamefont{S.~T.} \bibnamefont{Chu}},
  \bibinfo{journal}{IEEE Photon. Technol. Lett.} \textbf{\bibinfo{volume}{12}},
  \bibinfo{pages}{401} (\bibinfo{year}{2000}).

\bibitem[{\citenamefont{Almeida et~al.}(2004)\citenamefont{Almeida, Barrios,
  Panepucci, and Lipson}}]{AlmeidaNAT04}
\bibinfo{author}{\bibfnamefont{V.~R.} \bibnamefont{Almeida}},
  \bibinfo{author}{\bibfnamefont{C.~A.} \bibnamefont{Barrios}},
  \bibinfo{author}{\bibfnamefont{R.~R.} \bibnamefont{Panepucci}},
  \bibnamefont{and} \bibinfo{author}{\bibfnamefont{M.}~\bibnamefont{Lipson}},
  \bibinfo{journal}{\nat} \textbf{\bibinfo{volume}{431}}, \bibinfo{pages}{1081}
  (\bibinfo{year}{2004}).
\end{thebibliography}
\end{document}